\documentstyle[emulapj,epsfig]{article}

\begin{document}

\title{Propeller-driven  Outflows and Disk Oscillations}

\author{M.M.~Romanova}
\affil{Department of Astronomy, Cornell University, Ithaca, NY
14853-6801; ~ romanova@astro.cornell.edu}

\author{G.V.~Ustyugova}
\affil{Keldysh Institute of Applied Mathematics, Russian Academy of
Sciences, Moscow, Russia;~ ustyugg@spp.Keldysh.ru}

\author{A.V.~Koldoba}
\affil{Institute of Mathematical Modeling, Russian Academy of
Sciences, Moscow, Russia;~koldoba@spp.Keldysh.ru}

\author{R.V.E.~Lovelace}
\affil{Departments of Astronomy and Applied and Eng. Phys., Cornell
University, Ithaca, NY 14853-6801; ~RVL1@cornell.edu}

\medskip

\keywords{accretion, dipole --- magnetic fields --- stars: magnetic
fields --- X-rays: stars}

\begin{abstract}

      We report the discovery of propeller-driven outflows in axisymmetric
magnetohydrodynamic simulations of disk accretion to rapidly
rotating magnetized stars. Matter outflows  in a wide cone and is
centrifugally ejected from the inner regions of the disk.
     Closer to the axis there is
a strong, collimated, magnetically
dominated  outflow of energy and angular momentum
carried by the  open magnetic field lines from the star.
  The ``efficiency'' of the propeller may be
very high in the respect that most of the
incoming disk matter is expelled from the system in winds.
   The star spins-down rapidly
due to the magnetic interaction with the disk through closed
field lines and with corona through  open field lines.
      Diffusive and viscous interaction between magnetosphere and
the disk are important: no outflows were observed for very small
values of the diffusivity and viscosity.
      These simulation results are
applicable to the early stages of
evolution of  classical T Tauri
stars and to different stages of
evolution of cataclysmic variables and neutron
stars in binary systems.
      As an example, we have shown that young
rapidly rotating magnetized CTTSs  spin-down to their present slow
rotation in less than $10^6$ years.

\end{abstract}

\section{Introduction}

Different accreting magnetized stars are expected to be in the
propeller regime during their evolution.
      Examples include accretion
to fast rotating neutron stars (e.g., Davidson \& Ostriker 1973;
Illarionov \& Sunyaev 1975; Stella, White, \& Rosner 1986; Lipunov
1992; Treves, Colpi \& Lipunov 1993; Cui 1997; Alpar 2001; Mori \&
Ruderman 2003),
white dwarfs in cataclysmic variables, and classical
T Tauri stars  (CTTSs) at the
early stages of their evolution.
    The propeller regime is characterized by the fact that
the azimuthal velocity of the star's outer magnetosphere is larger
than the Keplerian velocity of the disk at that distance.

Different aspects of the propeller regime  were investigated
analytically (Davies, Fabian \& Pringle 1979;
Lovelace, Romanova \& Bisnovatyi-Kogan 1999; Ikhsanov 2002;
Rappaport, Fregeau, \& Spruit 2004; Eksi, Hernquist, \& Narayan
2005) and studied with computer simulations (Wang \& Robertson 1985;
Romanova et al. 2003; Romanova et al. 2004 -hereafter RUKL04).

However, only relatively ``weak" propellers were investigated
earlier (RUKL04), in which a star spins-down, but no significant
outflows were observed. In this paper we report on axisymmetric
(2.5D) simulations of ``strong" propellers, where significant part
of the disk matter is re-directed to the propeller-driven outflows.
       We observed that the disk oscillates between
``high" and ``low" states and expels matter to  conical outflows
quasi-periodically.
    The quasi-periodic outbursts associated with the disk-magnetosphere
interaction were predicted earlier by Aly \& Kuijpers (1990) and
observed in simulations by Goodson et al. (1997, 1999), Matt et al.
(2002), Romanova et al. (2002 - hereafter - RUKL02), Kato et al.
(2004), von Rekowski \& Brandenburg (2004), RUKL04, Yelenina \&
Ustyugova (2005). However, none of the earlier simulations
concentrated on the propeller stage, and only few oscillation
periods were obtained in earlier simulations. We report on modeling
of propeller stage, where numerous oscillations were observed.

\begin{figure*}[t]
\epsscale{2.}
  \plotone{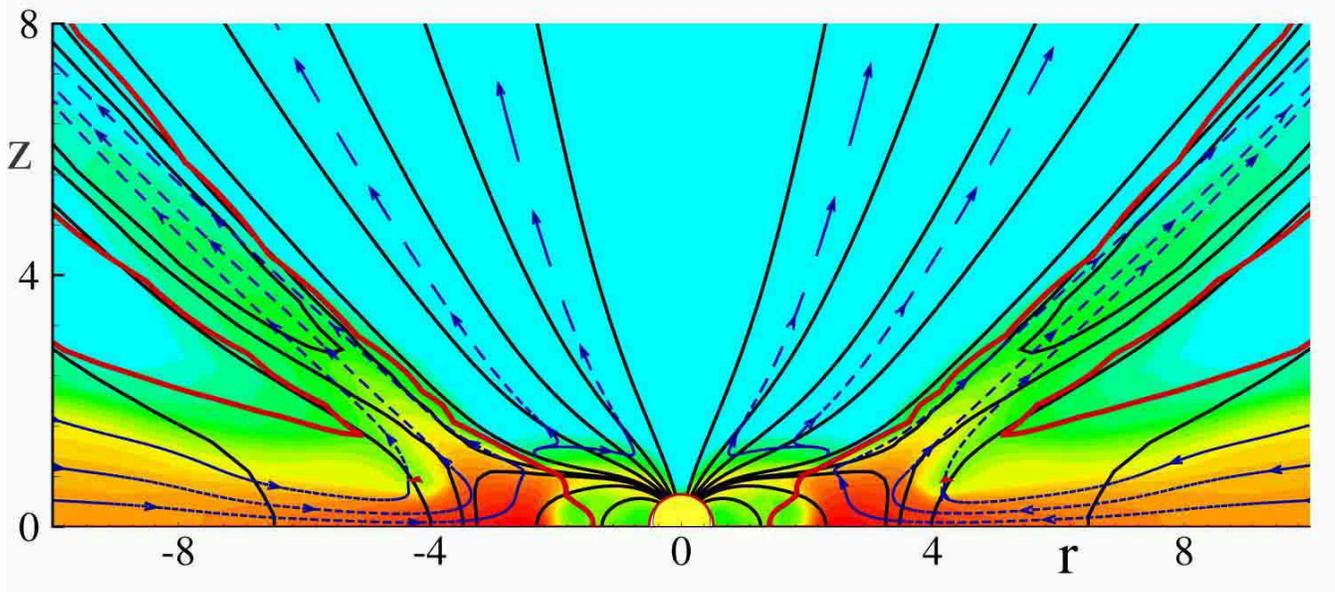}
\caption{Example of matter flow in the ``high" state (at
$t/P_0=924$). The color background shows the density distribution
which changes from red ($\rho=1.5$) to blue ($\rho=0.0005$). The
solid black lines are magnetic field lines. The dark-blue
lines/arrows are streamlines of the matter flow with the length of
arrows proportional to velocity. The bold red line shows the surface
where the matter pressure equals to magnetic pressure.}\label{Figure
1}
\end{figure*}

\begin{figure*}[t]
\epsscale{1.5} \plotone{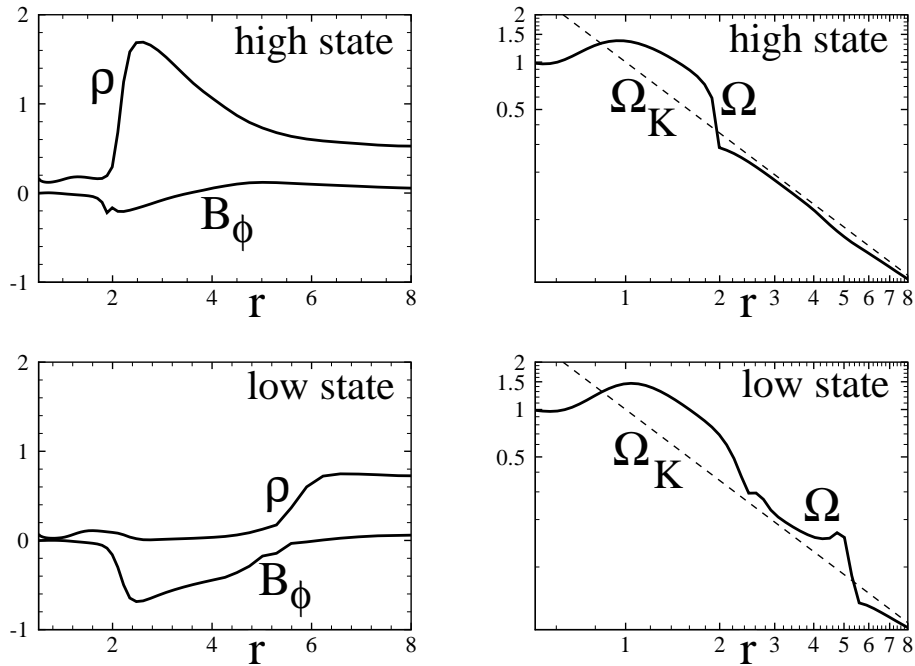} \caption{The left-hand panels show
the radial variation of the density $\rho$ and azimuthal magnetic
field $B_\phi$ above the equatorial plane. The right-hand panels
show the angular velocity $\Omega$ and the Keplerian angular
velocity $\Omega_K$ for comparison.} \label{Figure 2}
\end{figure*}

\begin{figure*}[b]
\epsscale{1.9}
  \plotone{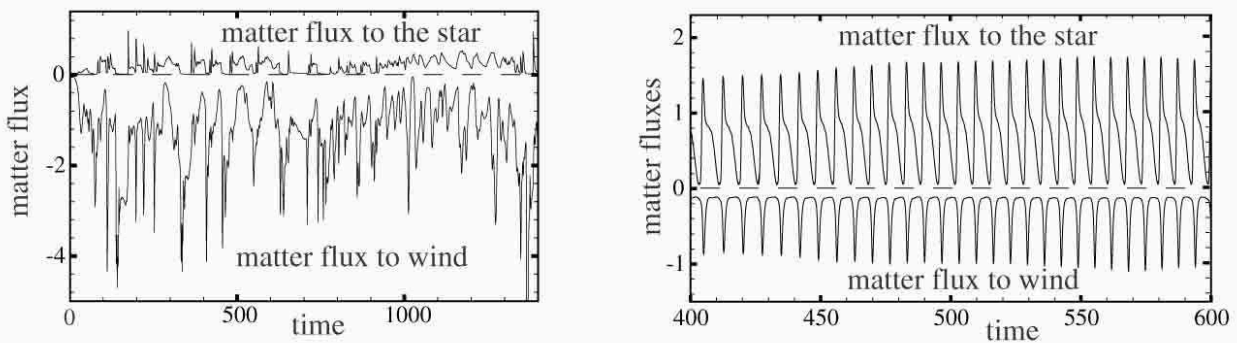}
\caption{The left-hand panel shows
the matter fluxes to the wind and to the star for our reference
case. The right-hand panel shows the quasi-periodic variations of
the mass fluxes which we find for larger viscosity, in this case,
$\alpha_{\rm v}=0.6$.}
\label{Figure 3}
\end{figure*}

\begin{figure*}[t]
\epsscale{1.3} \plotone{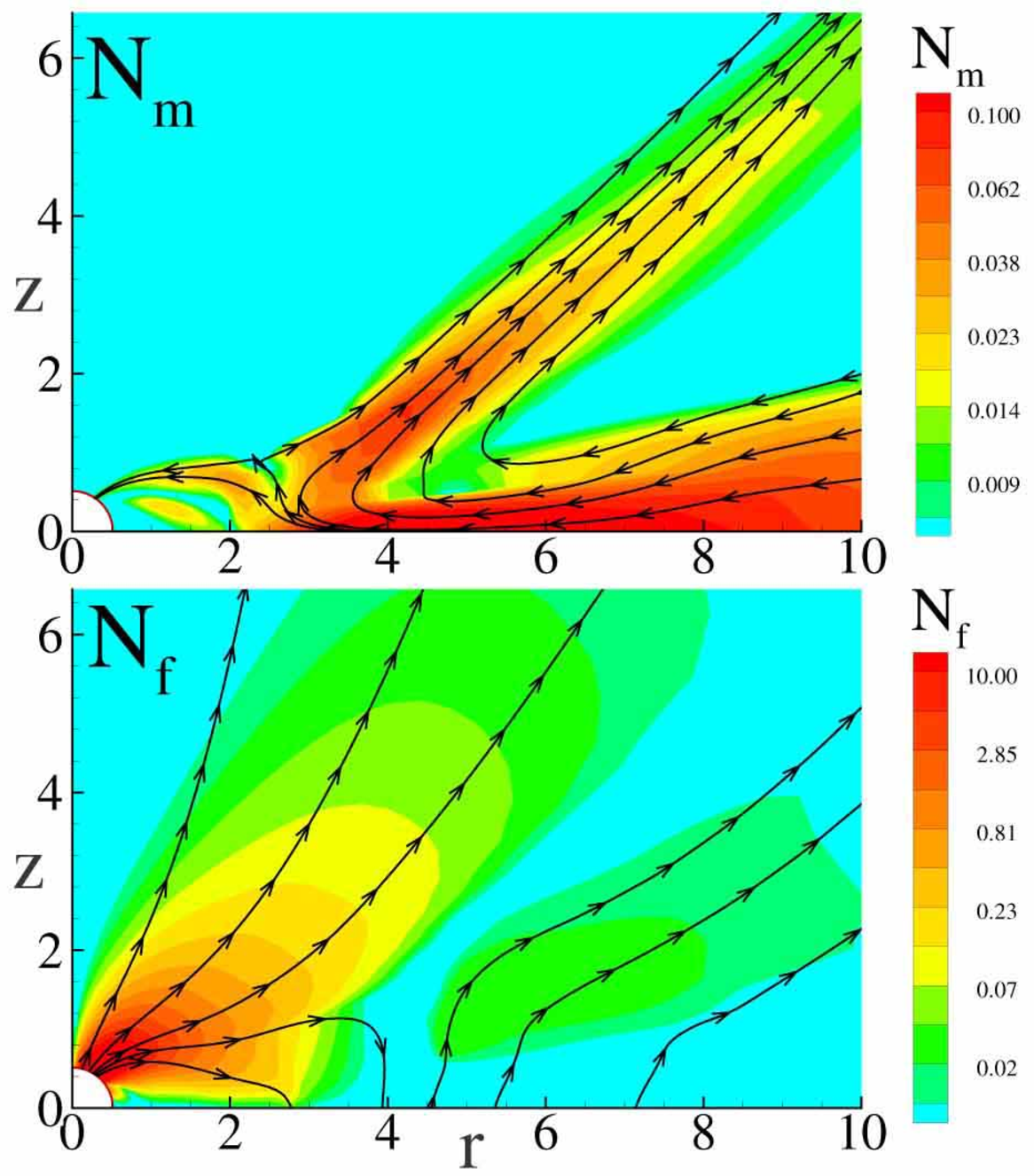} \caption{Color background and
streamlines show fluxes of angular momentum carried by the matter
(top panel) and by the field (bottom panel) in the ``high" state
($t/P_0=924$). Note that the scales are different, because $N_f$ is
very large near the surface of the star. However, the total
integrated fluxes $N_m$ and $N_f$ have comparable values.}
\label{Figure 4}
\end{figure*}

\section{Modeling of the Propeller-driven Outflows}

We have done axisymmetric MHD simulations of the interaction of an
accretion disk with magnetosphere of a rapidly rotating star.
      What
is meant by rapid rotation is that the corotation radius of the star
$r_{cr}=(GM_*/\Omega_*^2)^{1/3}$ is smaller than the magnetospheric
radius $r_m$ which is determined by the balance between the pressure
of the star's magnetic field and the ram pressure of the disk
matter.

     The numerical model we use is similar to that of RUKL02 and RUKL04.
   Specifically, (1) a spherical
coordinate system (r,$\theta$,$\phi$) is used  to give high
resolution near the dipole; (2) the complete set of MHD equations is
solved to find the eight variables ($\rho, v_r, v_\theta, v_\phi,
B_r, B_\theta, B_\phi, \varepsilon$) (with $\varepsilon$ the
specific internal energy); (3) a Godunov-type numerical method is
used;
   (4) special ``quiescent" initial
conditions were used so that we were able to observe slow viscous
accretion from beginning of simulations (see details in RUKL02).
   Compared to RUKL02, we now include the magnetic
diffusivity.
  The viscosity
and diffusivity are determined by turbulent fluctuations of the
velocity and magnetic field (e.g., Bisnovatyi-Kogan \& Ruzmaikin
1976) with both the kinematic viscosity $\nu_{\rm t}$ and the
magnetic diffusivity $\eta_{\rm t}$ of the
disk plasma described by
$\alpha-$coefficients as in the Shakura and Sunyaev model.
      That is,
$\nu_{\rm t}=\alpha_{\rm v} c_s^2/\Omega_K$ and $\eta_{\rm
t}=\alpha_{\rm d} c_s^2/\Omega_K$, where $\Omega_K$ is the Keplerian
angular velocity in the disk, $c_s$ is the isothermal sound speed,
and $\alpha_{\rm v}$ and $\alpha_{\rm d}$ are dimensionless
coefficients $\lesssim 1$.
    In RUKL04 we investigated a range of small viscosities and
diffusivities, $\alpha_{\rm v},~\alpha_{\rm d}\sim 0.01-0.02$ and
found no significant matter outflows.
    This paper  investigates a
wider range of $\alpha-$ parameters and finds
substantial outflows for
$\alpha_{\rm v} \gtrsim 0.1$ and $\alpha_{\rm d} \gtrsim 0.1$
in the propeller regime.
        Enhanced turbulence near the disk/magnetosphere
boundary may arise because the  radial gradient of the specific
angular momentum is negative resulting in instability (Ustyugova et
al. 2005).
   To model the diffusivity
terms in the MHD equations, we used an implicit numerical scheme and
the ICCG-method of solution of  linear equations.

{\bf 2.1.~Reference Units}.
    The MHD equations were solved in
dimensionless form so that the results
can be applied  to different systems.
    We take the reference mass $M_0 = M_*$, a scale $R_0=2 R_*$, and
a matter flux $\dot{M}_0$ which is close to the average matter flux
through the disk.
       Then we derive a reference density
$\rho_0=\dot M_0/{v_0 R_0^2}$,  a velocity $v_0=(GM_0/R_0)^{1/2}$, a
time-scale $t_0=R_0/v_0$, and an angular velocity $\Omega_0=1/t_0$.
The reference magnetic field is $B_0^2=\rho_0 v_0^2$.
   Thus $B_0=\sqrt{\rho_0} v_0$.
   A reference magnetic moment is $\mu_0=B_0
R_0^3 = \sqrt{\dot M_0 v_0} R_0^2$.
   The value of the  magnetic
moment used in our simulations $\mu_*$ is typically $10$ times larger
than $\mu_0$ so that we introduce new reference variable
$\mu_{00}=10\mu_0$.
   The magnetic field at the surface of the star is
$B_*=\mu_{00}/R_*^3$.
    The reference angular momentum flux is
$N_0=\dot M_0 v_0 R_0$.
   We measure time in units of the
rotational period of a Keplerian disk at $R_0$, $P_0=2\pi t_0$.
  We solve the  MHD equations for the normalized variables,
$\tilde\rho=\rho/\rho_0$, $\tilde v=v/v_0$,
$\tilde B= B/B_0$, etc. and below show  plots for normalized
variables (with tilda's dropped). In paragraph 2.5 we show an example
for CTTSs and millisecond pulsars in real units.

{\bf 2.2.~ Disk Oscillations and Outflows}.~
   Here, we discuss
results for a representative simulation run where outflows occurred.
The parameters are $\mu_*=\mu_{00}$, $\Omega_*=\Omega_0$
(co-rotation radius $r_{cr}=R_0$), $\alpha_{\rm v}=0.3$, and
$\alpha_{\rm d}=0.2$.
    The disk-magnetosphere interaction was found to be
quasi-periodic. The system oscillates  between a ``high" state,
where the inner radius of the disk is closest to the star, and  a
``low" state where the disk is at the largest distance from the
star.
    Figure 1 shows an example of matter flow in the ``high" state.
   The outflow is launched into a conical shell of half-angle $\theta\sim
45^\circ-60^\circ$.
    Many field lines are opened,
and a major part  of the matter flow to the wind  is along the
neutral line of the magnetic field. Analysis of the forces  shows
that the dominant force ``pushing" matter to the wind is the
centrifugal force (Blandford \& Payne 1982). At smaller $\theta$
there is a magnetically dominated outflow of the much lower-density
matter, which propagates with high-velocity along the open field
lines of the star (see, e.g., Lovelace et al. 2002; Ustyugova et al.
2005).

Our simulations show that: (1) matter accumulates in the disk and
moves inward; (2) it comes close to the star and penetrates
diffusively into the closed magnetosphere;
   (3) the disk matter acquires angular momentum
from the rapidly rotating magnetosphere  and is
expelled as outflows;
  (4) a small amount of matter accretes
to the star through a funnel flow;
  (4) the disk is pushed outward by
the rapidly rotating magnetosphere and the cycle repeats.
   Many field lines inflate and open
during outflow stage (see also Fendt \& Elstner 2000) and some
magnetic flux annihilates and may be a source of X-ray flares which
are often observed in young stars (Feigelson \& Montmerle 1999).

Figure 2  shows the radial distribution of density
$\rho$, angular velocity $\Omega$, and azimuthal magnetic field
$B_\phi$ in the ``high" state (top panels) and ``low" state (bottom
panels).
   In the ``high" state, the disk has a closest approach to the
star of $r_d\approx 2.5$, and the density in the inner disk is large.
   The magnetosphere rotates with super-Keplerian velocity. Azimuthal
component of the field is small.
   In the ``low" state, the disk is
pushed outward, the density is lower,
  and the magnetic field lines are twisted
up, forming modified expanded magnetosphere similar
to the  case of ``weak"
propellers (RUKL04). We conclude that no outflows were observed in
weak propellers (RUKL04), because the diffusivity and specific
matter flux were too low.

{\bf 2.3.~ Efficiency of Propeller and Variability}. We performed
simulations for a range of stellar magnetic moments $\mu_*$, angular
velocities $\Omega_*$, and viscosity and diffusivity coefficients,
$\alpha_{\rm v}$ and $\alpha_{\rm d}$, and calculated the
time-averaged matter fluxes to the star $\langle\dot M_{s}\rangle$
and to the wind, $\langle\dot M_{w}\rangle$, and the efficiency of the
propeller,
$$
{\cal R}\equiv \frac{\langle\dot M_{w}\rangle}{\langle\dot
M_{s}\rangle}\approx 13.0 ~\bigg(\frac{\mu_*}{\mu_{00}}\bigg)^2
\bigg(\frac{\Omega_*}{\Omega_0}\bigg)^{10.5} \bigg(\frac{\alpha_{\rm
d}}{0.2}\bigg)^{2.1} \bigg(\frac{\alpha_{\rm v}}{0.2}\bigg)^{2.5}.
$$
    The ratio ${\cal R}$
may be very large, ${\cal R} \gtrsim 50-100$; that is, almost all of
the matter coming inward in the disk may be ejected by the rapidly
rotating magnetosphere.
    The dependence of $\cal R$ on $\mu_*$,
$\Omega_*$, and $\alpha_{\rm v}$ is such that stronger ejection is
observed in cases of stronger field, faster rotation and larger
viscosity.
    However, the dependence on $\alpha_{\rm d}$ has a turnover
point at approximately $0.2$.
     For $\alpha_{\rm d} \lesssim 0.2$,
$\cal R$ increases with $\alpha_{\rm d}$, while for larger values it
decreases (Ustyugova et al. 2005).
    This paper shows the dependencies only for $\alpha_{\rm d} \lesssim 0.2$.

It is important to note that no significant outflows were observed
when the diffusivity was relatively low, $\alpha_{\rm d}\lesssim 0.1$.
   This is because at low  $\alpha_{\rm d}$
the penetration of the disk matter
into the  magnetosphere is not significant.
    Further, no outflows were observed at low viscosity.
Analysis of the stresses show  that the viscous stress is largest
at the disk-magnetosphere boundary, and thus it adds to the
``friction" and angular momentum transport from magnetosphere to the
disk.
    From other side, the matter flux in the disk is proportional
to the $\alpha_{\rm v}$ and at larger $\alpha_{\rm v}$, disk
penetrates to deeper, faster rotating layers of magnetosphere. Both
factors are important in generation of outflows, which appear at
$\alpha_{\rm v} \gtrsim 0.1$.
    Outflows were observed at a wide range of
the magnetic Prandtl numbers, $P_m=\nu_{\rm t}/ \eta_{\rm t} =
\alpha_{\rm v}/\alpha_{\rm m} \approx  0.2 - 6$ and thus do not
require the dominance of viscosity or diffusivity. Instead, both
$\alpha$ parameters should be larger than $0.1$.
   Note, that the observed
oscillations are completely determined by the processes at the
disk-magnetosphere boundary.
   They are different from the
viscous instability oscillations of the disk (Kato 1978)
which can not be investigated by our numerical model.

The amplitude of the fluxes changes rapidly (see Figure 3).
    There is a
typical time-scale of variations $\tau_{\rm qpo}$ in each case which
depends on main parameters $\mu_*$, $\Omega_*$, $\alpha_{\rm d}$,
and $\alpha_{\rm v}$.
    For given
$\alpha_{\rm d}$ and $\alpha_{\rm v}$, the ``quasi-period"
increases with  $\mu_*$ and $\Omega_*$.
    It varies
in the range $\tau_{\rm qpo}=(5-100) P_0$.
    We observed that for  values of the diffusivity
$\alpha_{\rm d} =0.2$, but relatively high viscosities $\alpha_{\rm
v}$ ranging from $0.6$ to $1$, the oscillations become highly
periodic. In one of of sample runs  a quasi-period changes from
$\tau_{qpo}=10 P_0$ to $6.5 P_0$ (see Figure 3, right panel). Period
changes due to the fact that  the inner disk radius moved closer to
the star.

{\bf 2.4. Angular Momentum Transport and  Spinning-down}. The
angular momentum flux carried by the disk matter is re-directed by
the rapidly rotating magnetosphere to the outflows (top panel of
Figure 4).
      Furthermore, there is a strong outflow of angular
momentum (and energy) carried by the twisted open magnetic field
lines from the star  $\langle N_f\rangle$, the Poynting flux, and
the closed field lines connecting the star and disk (bottom panel of
Figure 4; see also Lovelace et al. 2002).
    The time averaged total angular
momentum flux  from the star is $ \langle N_s \rangle \approx - 3.1
({\mu_*}/{\mu_{00}})^{1.1} ({\Omega_*}/{\Omega_0})^{2.0}
({\alpha_{\rm d}}/{0.2})^{0.46} ({\alpha_{\rm v}}/{0.2})^{0.1}. $
  For our typical  parameters the
spin-down associated with open and closed field lines are
comparable.
   However, at larger $\Omega_*$ and/or $\bf{\mu}_*$, the
outflow along the open field lines dominates (see related cases in
Lovelace et al. 1995;
     Matt \& Pudritz 2004), while at
lower $\Omega_*$ or $\bf{\mu}_*$,
the situation reverses and a larger
flux is associated with the closed field lines (like in Ghosh \&
Lamb 1979; RUKL02).

The  spin-down time-scale follows from equating  the torque
${\langle N_s\rangle} N_0$ to $I_* d\Omega_*/dt$, where $I_*\approx
10^{45}{\rm gcm}^2$ is the moment of inertia of the star.
     For the period of the star
$P_*=2\pi/\Omega_*$, we obtain: $P_*(t) =P_*(0)(1+t/t_{\rm sd})$,
where the spin-down time is
$$ t_{\rm sd}\approx 0.036 \bigg(\frac{M_*}{\dot M_0}\bigg)
\bigg(\frac{\mu_{00}}{\mu_*}\bigg)^{1.1}
 \bigg(\frac{0.2}{\alpha_{\rm d}}\bigg)^{0.46}
\bigg(\frac{0.2}{\alpha_{\rm v}}\bigg)^{0.1} . $$

{\bf 2.5. Young CTTSs and Millisecond Pulsars}.
   Our results
can be directly applied to stars with relatively small
magnetospheres, $r_m \lesssim (3-10) R_*$, for example, to CTTSs or
to accreting millisecond pulsars.  Thus, for a   CTTS with a  mass
$M_*=0.8 M_\odot$, radius $R_*=2 R_\odot$, an accretion rate $\dot
M_0\approx 5\times 10^{-8}~{\rm M_\odot/yr}$, and other typical
parameters $\alpha_{\rm d}=0.2$, $\alpha_{\rm v}=0.2$, and
$\mu_*=\mu_{00}$, we obtain $t_{\rm sd}\approx 5.8\times 10^5~{\rm
yr}$.
   We also derive dimensional values:
$\mu_*\approx 6.2\times10^{36}~
{\rm G cm^3}$, $B_*\approx2.2\times
10^3~{\rm G}$ and $P_*\approx 1~{\rm day}$.
   Thus, if young CTTSs
have strong magnetic field, then they  spin-down
rapidly to their
presently observed slow rotation rate.

{\it Accreting millisecond pulsars} have a different history of
evolution: they spin-up from slow to fast rotation.
   However, they
may have episodes in the propeller regime.
    For a neutron star with
mass $M_*=1.4 M_\odot$ and similar accretion rate $\dot M_0\approx
5\times 10^{-8}~{\rm M_\odot/yr}$, we find
$t_{sd}\approx10^6~{\rm yr}$.
   Taking $R_*=10^6~{\rm cm}$ and
$\mu_*=\mu_{00}$, we find $\mu_*\approx7.0\times
10^{27}~{\rm G cm^3}$, $B_*\approx7\times 10^9~\rm G$,
$P_*\approx1.3~{\rm ms}$.
   In this case $t_{sd}$ represents the
time-scale of spin change for rapidly rotating millisecond
pulsars.

\section{Conclusions}

    In the propeller regime of disk
accretion to a rapidly rotating star, we find
from axisymmetric MHD simulations
that the disk oscillates strongly
and gives quasi-periodic  outflows
of matter to
wide-angle ($\chi\approx 45^\circ - 60^\circ$) conical winds.
    At the same time there is strong field-dominated (or Poynting)
outflow of energy, angular momentum and matter along the open field
lines extending from the poles of the star. Matter outflows with
high velocities and is magnetically driven.
     The outflows occur for conditions where
the magnetic diffusivity and viscosity
are significant, $\alpha_{\rm v,d}\gtrsim 0.1$.
       For smaller values of the diffusivity, the disk
oscillates but no outflows are observed (RUKL04).
   The observed
oscillations and outbursts are a robust result, based on a numerous
simulations at different parameters with more than a $100$
oscillation periods observed in many runs.
  The period of oscillations varies in different runs in
the range $\tau_{\rm qpo}\sim (5-100) P_*$.
  It increases with
$\mu_*$ and $\Omega_*$.
    We observed that the oscillations for
relatively large $\alpha_{\rm v}$
become highly periodic with definite quasi-periods.
    More detailed analysis of
these features will be reported later. A  star spins-down rapidly
due to both the disk-magnetosphere interaction and the angular
momentum outflow along the open field lines.
   The results are applicable to young CTTSs,
neutron stars, and cataclysmic
variables.

\acknowledgments This work was supported in part by NASA grants
NAG5-13220, NAG5-13060, NNG05GG77G and by NSF grants AST-0307817,
AST-0507760. AVK and GVU were partially supported by RFBR
03-02-16548 grant. The authors thank  D. Proga, D.
Rothstein, and C. Fendt for stimulating discussions.


\begin{references}


\reference{alp01} Alpar, M.A. 2001, ApJ, 554, 1245

\reference{aly90} Aly, J.J., \& Kuijpers, J. 1990, A\&A, 227, 473 73

\reference{bis76} Bisnovatyi-Kogan, G.S., \& Ruzmaikin, A.A. 1976,
AP\&SS, 42, 401


\reference{bla82} Blandford, R.D., \& Payne, D.G. 1982, MNRAS, 199,
883

\reference{} Cui, W. 1997, ApJ, 482, L163

\reference{dav73} Davidson, K., \& Ostriker, J.P. 1973, ApJ, 179,
585

\reference{} Davies, R.E., Fabian, A.C., \& Pringle, J.E. 1979,
MNRAS, 186, 779




\reference{} Eksi, K. Y., Hernquist, L., \& Narayan, R. 2005,
ApJ, 623, L41-L44



\reference{} Feigelson, E.D., \& Montmerle, T. 2005, Ann. Rev.
Astron. \& Astrophys., 37, 363

\reference{fen02} Fendt, C., \& Elstner, D. 2002, A\&A, 363, 208

\reference{gho79} Ghosh, P., \& Lamb, F.K. 1979, ApJ, 234, 296

\reference{goo99} Goodson, A.P., B\"ohm, K.-H., Winglee, R. M. 1999,
ApJ, 524, 142


\reference{goo99} Goodson, A.P., Winglee, R. M., \& B\"ohm, K.-H.,
1997, ApJ, 489, 199

\reference{ill75} Illarionov, A.F., \& Sunyaev, R.A. 1975, A\&A,
39, 185

\reference{ikh} Ikhsanov, N.R. 2002, A\&A, 381, L61

\reference{kat78} Kato, S. 1978, MNRAS, 185,,629


\reference{} Kato, Y., Hayashi, M.R., \& Matsumoto, R. 2004, ApJ,
600, 338


\reference{lip92} Lipunov, V.M. 1992, {\it Astrophysics of Neutron
Stars}, (Berlin: Springer Verlag)



\reference{lov95} Lovelace, R.V.E., Romanova, M.M., \&
Bisnovatyi-Kogan, G.S. 1995, MNRAS, 275, 244


\reference{lov99} Lovelace, R.V.E., Romanova, M.M., \&
Bisnovatyi-Kogan, G.S.
1999, ApJ, 514, 368

\reference{lov02} Lovelace, R.V.E., Li, H., Koldoba, A.V.,
Ustyugova, G.V., \& Romanova, M.M. 2002, ApJ, 572, 445

\reference{} Matt, S., Goodson, A.P., Winglee, R.M., \& B\"ohm,
K.-H. 2002, ApJ, 574, 232


\reference{} Matt, S., \& Pudritz, R.E. 2004, ApJ, 607, L43


\reference{mor03} Mori, K., \& Ruderman, M.A. 2003, ApJ 592, L75


\reference{rap04} Rappaport, S. A., Fregeau, J. M., \& Spruit, H.
2004, 606, 436

\reference{} Romanova, M.M., Toropina, O.D., Toropin, Yu.M., \&
Lovelace, R.V.E. 2003, ApJ, 588, 400

\reference{Rom02} Romanova, M.M., Ustyugova, G.V., Koldoba, A.V., \&
Lovelace, R.V.E. 2002, ApJ, 578, 420 (RUKL02)


\reference{Rom04} Romanova, M.M., Ustyugova, G.V., Koldoba, A.V., \&
Lovelace, R.V.E. 2004, ApJ, 616, L151 (RUKL04)

\reference{} Stella, L., White, N.E., \& Rosner, R. 1986, ApJ,
308, 669

\reference{} Treves, A., Colpi, M., \& Lipunov, V.M. 1993, A\&A,
269, 319

\reference{} von Rekowski, B., \& Brandenburg, A. 2004, A \& A, 420,
17

\reference{} Ustyugova, G.V., Koldoba, A.V., Romanova, M.M., \&
Lovelace, R.V.E. 2005, ApJ, in press


\reference{} Wang, Y.-M., \& Robertson, J.A. 1985, A\&A 151, 361

\reference{} Yelenina, T.G., \& Ustyugova, G.V. 2005, preprint of
the Keldysh Institute of Applied Mathematics
(http://www.keldysh.ru/papers/2005/prep16/prep2005$\underline~$16en.pdf)



\end{references}
\end{document}